\documentclass[twocolumn,letter]{jpsj3}
\usepackage{graphicx}
\usepackage{bm}
\usepackage{cite}
\usepackage{color}
\usepackage{amssymb}
\usepackage{amsmath}
\usepackage{revsymb}
\usepackage{here}

\title{Low-Energy Excitation Spectra in the Excitonic Phase of Cobalt Oxides}

\author{
Tomoki Yamaguchi$^1$\thanks{t.yamaguchi@chiba-u.jp},
Koudai Sugimoto$^2$, and Yukinori Ohta$^1$}
\inst{
$^1$Department of Physics, Chiba University, Chiba 263-8522, Japan \\
$^2$Center for Frontier Science, Chiba University, Chiba 263-8522, Japan \\} %\\

\date{\today}

\abst{
We study the excitonic phase and low-energy excitation spectra of perovskite cobalt oxides.  
Constructing the five-orbital Hubbard model defined on the three-dimensional cubic lattice for the $3d$ 
bands of Pr$_{0.5}$Ca$_{0.5}$CoO$_3$, we calculate the excitonic susceptibility in the normal state 
in the random-phase approximation (RPA) to show the presence of the instability toward excitonic 
condensation.  On the basis of the excitonic ground state with a magnetic multipole obtained in the 
mean-field approximation, we calculate the dynamical susceptibility of the excitonic phase in the 
RPA and find that there appear a gapless collective excitation in the spin-transverse mode 
(Goldstone mode) and a gapful collective excitation in the spin-longitudinal mode (Higgs mode).  
The experimental relevance of our results is discussed. 
}

\begin{document}
\maketitle

%\section{Introduction} %[[[

The Bose--Einstein condensation of fermion pairs is one of the most intriguing phenomena 
in condensed matter physics.  The excitonic phase (EP) is representative of such a pair 
condensation,\cite{Jerome1967PR, Halperin1968RMP, Littlewood2004JPCM, Kunes2015JPCM} 
where holes in valence bands and electrons in conduction bands spontaneously form pairs 
owing to attractive Coulomb interaction.  After Mott's prediction of the EP half a century 
ago,\cite{Mott1961PM} a number of candidate materials for this phase have come to 
our attention.  Among them are the transition-metal chalcogenides 
1$T$-TiSe$_2$\cite{DiSalvo1976PRB, Cercellier2007PRL, Monney2009PRL} and 
Ta$_2$NiSe$_5$,\cite{Wakisaka2009PRL, Kaneko2013PRB, Seki2014PRB} where the electrons and 
holes on different atoms are considered to form spin-singlet pairs to condense into the EP, which is 
accompanied by lattice distortion.\cite{Kaneko2015PRB}

Another class of materials includes the perovskite cobalt oxides,\cite{Kunes2014PRB, Nasu2016PRB, Afonso2016} 
where the valence-band holes and conduction-band electrons form spin-triplet pairs in different 
orbitals on the same atoms.  Pr$_{0.5}$Ca$_{0.5}$CoO$_3$ (PCCO) is an example in which the 
``metal-insulator'' phase transition is observed at $T_c\simeq 80$ K, which is associated with 
a sharp peak in the temperature dependence of the specific heat and a drop in the magnetic 
susceptibility below $T_c$,\cite{Tsubouchi2002PRB} together with a valence transition of Pr 
ions.\cite{Hejtmanek2010PRB,Garcia-Munoz2011PRB}  
Some results of experiments indicate that the resistivity is in fact small and nearly temperature 
independent below $T_c$,\cite{Hejtmanek2013EPJB} suggesting that the bands may not be fully 
gapped.  Note that no local magnetic moments are observed, but the exchange splitting of the 
Pr$^{4+}$ Kramers doublet occurs,\cite{Hejtmanek2013EPJB} the result of which may therefore be 
termed as a {\em hidden} order, and also that no clear signatures of the spin-state transition are 
observed in the X-ray absorption spectra.\cite{Herrero-Martin2012PRB, Hejtmanek2013EPJB}

Kune\v{s} and Augustinsk\v{y} argued that the anomalies of PCCO can be attributed to the EP 
transition,\cite{Kunes2014PRB} whereby they applied the dynamical-mean-field-theory calculation 
to the two-orbital Hubbard model defined on a two-dimensional square lattice and claimed that 
the anomalous behaviors of the specific heat, dc conductivity, and spin susceptibility can 
be explained.  They also performed the LDA$+U$ band-structure calculation and showed that 
the magnetic multipole ordering occurs in PCCO as a result of the excitonic condensation.  
LaCoO$_3$ under a high magnetic field is another example of the possible realization of the 
EP,\cite{Ikeda2016PRB} which was substantiated by the theoretical calculations based on the 
two-orbital Hubbard and related models in two-dimension.\cite{Sotnikov2016SR, Tatsuno2016JPSJ}  
In these materials, cobalt ions are basically in the Co$^{3+}$ valence state with a $3d^6$ 
configuration, where the three $t_{2g}$ orbitals are mostly filled with electrons and the two $e_g$ 
orbitals are nearly empty.  The low-spin state is thus favorable for the condensation of excitons. 
%that are formed between holes in the three $t_{2g}$ bands and electrons in the two $e_g$ bands.  

In this work, motivated by the above development in the field, we will study the EP of PCCO 
using a realistic Hubbard model, taking into account all five $3d$ orbitals of Co ions 
arranged in the three-dimensional cubic lattice of the perovskite structure.  
The noninteracting tight-binding bands are determined from first principles and the electron-electron 
interactions in the $3d$ orbitals are fully taken into account in each Co ion.  
We will then study the excitonic fluctuations in the normal state via the calculation of the 
excitonic susceptibility in the random phase approximation (RPA) and show that the instability 
toward the EP actually occurs in this model.  The ground state of this model is then calculated 
in the mean-field approximation, whereby we find that the EP with a magnetic multipole order 
actually occurs.  We will also calculate the dynamical susceptibility of both spin-transverse 
and spin-longitudinal modes in the EP to clarify the presence of the gapless Goldstone and gapful 
Higgs modes in the excitation spectra.  The experimental relevance of our results will be discussed.  

%]]]

%\section{MODEL AND METHOD} %[[[

\begin{figure}[thb]
\begin{center}
\includegraphics[width=8.5cm]{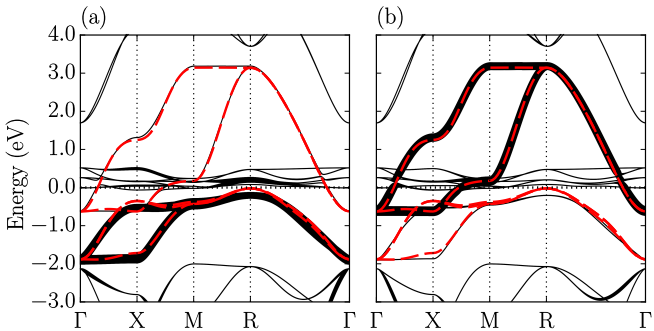}
\caption{(Color online) 
Calculated band dispersions of PrCoO$_3$ plotted along the lines connecting 
$\Gamma(0,0,0)$, X$(\pi, 0, 0)$, M$(\pi, \pi, 0)$, and R$(\pi, \pi, \pi)$. 
The width of the black solid curves is in proportion to the weight of 
(a) the Co $t_{2g}$ orbitals and (b) the Co $e_g$ orbitals.  
The tight-binding band dispersions obtained using the maximally localized 
Wannier functions are also shown by red dashed curves.  
}\label{fig:band}
\end{center}
\end{figure}

%\subsection{Band structure of PrCoO3}

The crystal structure of PCCO belongs to the $P_{nma}$ space group, where the CoO$_6$ 
octahedra are rotated and the cubic perovskite structure is distorted with two independent 
Co ions in the unit cell,\cite{Tsubouchi2002PRB} giving rise to complexity in the analysis 
of the EP in PCCO.  We instead make use of the crystal structure of PrCoO$_3$, which is 
a perfect cubic perovskite with the lattice constant $a=3.82$ \AA \cite{Wei-RanCPL2005} 
(hereafter taken as the unit of length).  The electronic structure is then calculated from 
first principles using the WIEN2k code~\cite{WIEN2k}.
The obtained band dispersions are illustrated in Fig.~\ref{fig:band}, where we find that the 
bands near the Fermi energy come from the Pr $4f$ orbitals (giving narrow dispersions) 
and Co $3d$ orbitals (giving wide dispersions).  For the latter, we find that the valence 
bands come from the $t_{2g}$ manifold of Co $3d$ orbitals and the conduction bands 
come from the $e_g$ manifold, as indicated in the weight plot.  We, moreover, find that 
the $t_{2g}$ bands and $e_g$ bands are orthogonal to each other without hybridization, 
providing us with an ideal situation for excitonic condensation.  
We also performed the band-structure calculation of PCCO, arranging the Pr and Ca ions 
regularly, and confirmed that the bands remain qualitatively unchanged, supporting 
the validity of the rigid-band approximation.  
Note that PCCO contains both the Co$^{3+}$ and Co$^{4+}$ ions,\cite{Hejtmanek2013EPJB} 
while PrCoO$_3$ contains only the Co$^{3+}$ ions and shows no signatures of the phase 
transition.\cite{Pandey2008PRB}  The change in the valence state of Co ions, which may 
lead to a better nesting feature of the Fermi surfaces, seems to play an important role in 
the EP transition.  
Hereafter, we focus on the $3d$ bands of the Co ions, and assuming that the $4f$ bands 
of Pr ions act as a bath of electrons and, together with the presence of Ca$^{2+}$ ions, 
the valence state of Co ions is kept to be in the $3d^6$ configuration for 
simplicity\cite{Kunes2014PRB} unless otherwise stated.  

%\subsection{Multiorbital Hubbard Model}

Let us now set up the Hamiltonian $H=H_0+H_{\rm int}$ for the modeling of the $3d$ electrons 
of PCCO.  The kinetic energy term $H_0$ is defined in the tight-binding approximation as 
\begin{equation}
  H_{0} = \sum_{i, \mu, \sigma} \epsilon_{\mu} c^{\dagger}_{i,\mu,\sigma} c_{j,\mu,\sigma}
	+  \sum_{i, j} \sum_{\mu,\nu} \sum_{\sigma} t_{ij,\mu\nu} c^{\dagger}_{i,\mu,\sigma} c_{j,\nu,\sigma},
\end{equation}
where $c^{\dagger}_{i,\mu,\sigma}$ is the creation operator of a spin-$\sigma$ $(=\uparrow,\downarrow)$ 
electron on the orbital $\mu$ at site $i$, $\epsilon_{\mu} $ is the on-site energy of orbital 
$\mu$, and $t_{ij, \mu\nu}$ is the hopping integral between the orbital $\nu$ at site $j$ 
and the orbital $\mu$ at site $i$.  The orbitals $\mu$ and $\nu$ are labeled as 1 ($d_{xy}$), 
2 ($d_{yz}$), 3 $(d_{zx}$), 4 ($d_{x^2-y^2}$), and 5 ($d_{3z^2 - r^2}$).  
The 12 molecular orbitals for the $3d$ and $4f$ bands are obtained as the maximally localized 
Wannier functions,\cite{Kunes2010CPC,Mostofi2014CPC} thereby retaining only the $3d$ bands 
to determine the on-site energies and hopping integrals (up to 6th neighbors).  
The tight-binding band dispersions thus calculated reproduce the first-principles band structure 
well, as shown in Fig.~\ref{fig:band}.  
%Note that there are no hopping integrals between the $t_{2g}$ and $e_{g}$ orbitals.  

The on-site interaction term is defined as 
\begin{align}
  H_{\rm int}
	&= \dfrac{U}{2}\sum_{i,\mu,\sigma}
  		c^{\dagger}_{i\mu\sigma} c_{i\mu\sigma} c^{\dagger}_{i\mu-\sigma}c_{i\mu-\sigma}
\notag \\
	&+ \dfrac{U'}{2} \sum_{i,\sigma,\sigma'} \sum_{\mu\neq\nu}
  		c^{\dagger}_{i,\mu,\sigma} c_{i,\mu,\sigma} c^{\dagger}_{i,\nu,\sigma'} c_{i,\nu,\sigma'}
\notag \\
	&- \dfrac{J}{2} \sum_{i,\sigma,\sigma'} \sum_{\mu\neq\nu}
  		c^{\dagger}_{i,\mu,\sigma} c_{i,\mu,\sigma'} c^{\dagger}_{i,\nu,\sigma'} c_{i,\nu,\sigma}
\notag \\
	&+ \dfrac{J'}{2} \sum_{i,\sigma}\sum_{\mu\neq\nu}
  		c^{\dagger}_{i,\mu,\sigma} c_{i,\nu,-\sigma} c^{\dagger}_{i,\mu,-\sigma} c_{i,\nu,\sigma},
\label{eq:H_int}
\end{align}
where $U, U', J$, and $J'$ are the intraorbital Coulomb interaction, interorbital Coulomb interaction, 
Hund's rule coupling, and pair-hopping interaction, respectively.  
We assume the atomic-limit relations $U' = U - 2J$ and $J' = J$ for the interaction strengths, 
and we fix the ratio $J/U$ at $0.1$ in the present calculations.  

We apply the mean-field approximation to the interaction terms.  We assume the spin-triplet 
excitonic order in the presence of Hund's rule coupling \cite{Kaneko2014PRB} and write 
the order parameters as 
\begin{equation}
 \Delta_{\mu,\nu}
	= \sum_{\bm{k}, \sigma}  \sigma \left\langle c^{\dagger}_{\bm{k}+\bm{Q}, \mu, \sigma} c_{\bm{k}, \nu, \sigma} \right\rangle , 
\end{equation}
where $c_{\bm{k}, \mu, \sigma}$ is the Fourier component of $c_{i,\mu,\sigma}$ with the wave 
vector $\bm{k}$, and $\bm{Q}$ is an ordering vector.  
%$\sigma = 1$ $(-1)$ corresponds to up (down) spin. 
Note that when $\mu$ ($\nu$) is one of the $e_{g}$ orbitals, $\nu$ ($\mu$) is one of the $t_{2g}$ orbitals.  
All the terms irrelevant to this excitonic ordering are neglected for simplicity.  
We thus obtain the diagonalized mean-field Hamiltonian, 
\begin{equation}
 H^{\rm MF}
	= \sum_{\bm{k}_0, \epsilon, \sigma} E_{\bm{k}_0, \epsilon, \sigma}
		\gamma^{\dagger}_{\bm{k}_0, \epsilon, \sigma} \gamma_{\bm{k}_0, \epsilon, \sigma} , 
\end{equation}
where $\gamma_{\bm{k}_0, \epsilon, \sigma}$ is the canonical transformation of $c_{\bm{k}, \mu, \sigma}$ 
satisfying 
$c_{\bm{k}, \mu, \sigma} = \sum_{\epsilon} \psi_{\mu, m; \epsilon} (\bm{k}_0, \sigma) \gamma_{\bm{k}_0, \epsilon, \sigma}$ 
and $\epsilon$ is the band index.  
Since the excitonic order enlarges the unit cell, we write the wave vector as $\bm{k} = \bm{k}_0 + m \bm{Q}$, 
where $\bm{k}_0$ is the wave vector in the reduced Brillouin zone and $m$ is an integer.  
We carry out the summation with respect to $\bm{k}_0$ using the $50 \times 50 \times 50$ meshes 
in the reduced Brillouin zone.  
%The chemical potential is determined so as to fix the numbers of electrons as 6 per site.  

%\subsection{Dynamical Susceptibility}

We define the dynamical susceptibility as
\begin{multline}
\chi_{\substack{\lambda \mu \\ \kappa \nu}}^{s s'} (\bm{q}, \bm{q}', \omega)
 = \frac{i}{N} \sum_{\bm{k}, \bm{k}'} \int^\infty_0 d t \,  e^{i \omega t}
\\
	\times \langle  [ c_{\bm{k}, \kappa, \sigma_1}^\dagger(t)   c_{\bm{k} + \bm{q}, \lambda, \sigma_2} (t),
		c_{\bm{k}' + \bm{q}', \mu, \sigma'_1}^\dagger  c_{\bm{k}', \nu, \sigma'_2} ] \rangle,
\label{eq:Chi}
\end{multline}
where $N$ is the number of ${\bm k}$ points used, 
$c_{\bm{k}, \mu, \sigma}(t)$ is the Heisenberg representation of $c_{\bm{k}, \mu, \sigma}$, 
and $s$ denotes a spin pair $(\sigma_1, \sigma_2)$, taking the values 
$\uparrow$, $\downarrow$, $+$, and $-$ for $(\uparrow, \uparrow)$, $(\downarrow, \downarrow)$, 
$(\uparrow, \downarrow)$, and $(\downarrow, \uparrow)$, respectively.  
We write Eq.~\eqref{eq:Chi} as $\chi(\bm{q},\omega)$ when $\bm{q}=\bm{q}'$.  
The bare susceptibility is given by 
\begin{align}
 &{\chi_0^{ss'}}_{\substack{\lambda \mu \\ \kappa \nu}} (\bm{q}, \bm{q}+l\bm{Q}, \omega)
\notag \\
	=& -\frac{1}{N} \sum_{\bm{p_0}, m, n, \epsilon, \epsilon'}
		\frac{ f(E_{\bm{p}_0+\bm{q}, \epsilon, \sigma_1}) - f(E_{\bm{p}_0, \epsilon', \sigma_2})}
			{ E_{\bm{p}_0+\bm{q}, \epsilon, \sigma_1} - E_{\bm{p}_0, \epsilon', \sigma_2} - ( \omega + i \eta)} \notag \\
	& \times
		\psi_{\lambda, m ; \epsilon} (\bm{p}_0 + \bm{q}, \sigma_1)
		\psi_{\mu,m + n + l ; \epsilon}^\ast (\bm{p}_0 + \bm{q}, \sigma_2') \notag \\
	& \times
		\psi_{\kappa,m ; \epsilon'}^\ast (\bm{p}_0, \sigma_2)
		\psi_{\nu, m+n ; \epsilon'} (\bm{p}_0, \sigma_1') \delta_{\sigma_1, \sigma_2'} \delta_{\sigma_1', \sigma_2},
\end{align}
where the summation with respect to $\bm{p}_0$ runs over the reduced Brillouin zone.  
We set $\eta = 0.01$ eV.

\begin{table}[t]
\caption{
Nonzero elements of $V_{\substack{\mu \lambda \\ \nu \kappa}}^{ss'}$, where $s = (\sigma_1, \sigma_2)$ and $s' = (\sigma_1', \sigma_2')$.
}
\begin{tabular}{l|ccc}
\hline
 &
\shortstack[l]{$ \sigma_1 = \sigma_2 $ \\ $= \sigma'_1 = \sigma'_2$}&
\shortstack[l]{$\sigma_1 = \sigma_2 $ \\ $\neq \sigma'_1 = \sigma'_2$} &
\shortstack[l]{$\sigma_1 = \sigma_2' $ \\ $\neq \sigma_2 = \sigma_1'$}\\
\hline
$\mu = \nu = \kappa = \lambda$ & --  & $-U$ & $U$\\
$\mu = \kappa \neq \nu = \lambda$ & -- & $-J$ & $J$\\
$\mu = \nu \neq \kappa = \lambda$ & $-U + 3J$ & $-U+2J$ & $J$\\
$\mu = \lambda \neq \nu = \kappa$ & $U-3J$ & $-J$ & $U-2J$\\
\hline
\end{tabular}
\label{tb:interaction}
\end{table}

\begin{figure}[h]
\begin{center}
\includegraphics[width=\linewidth]{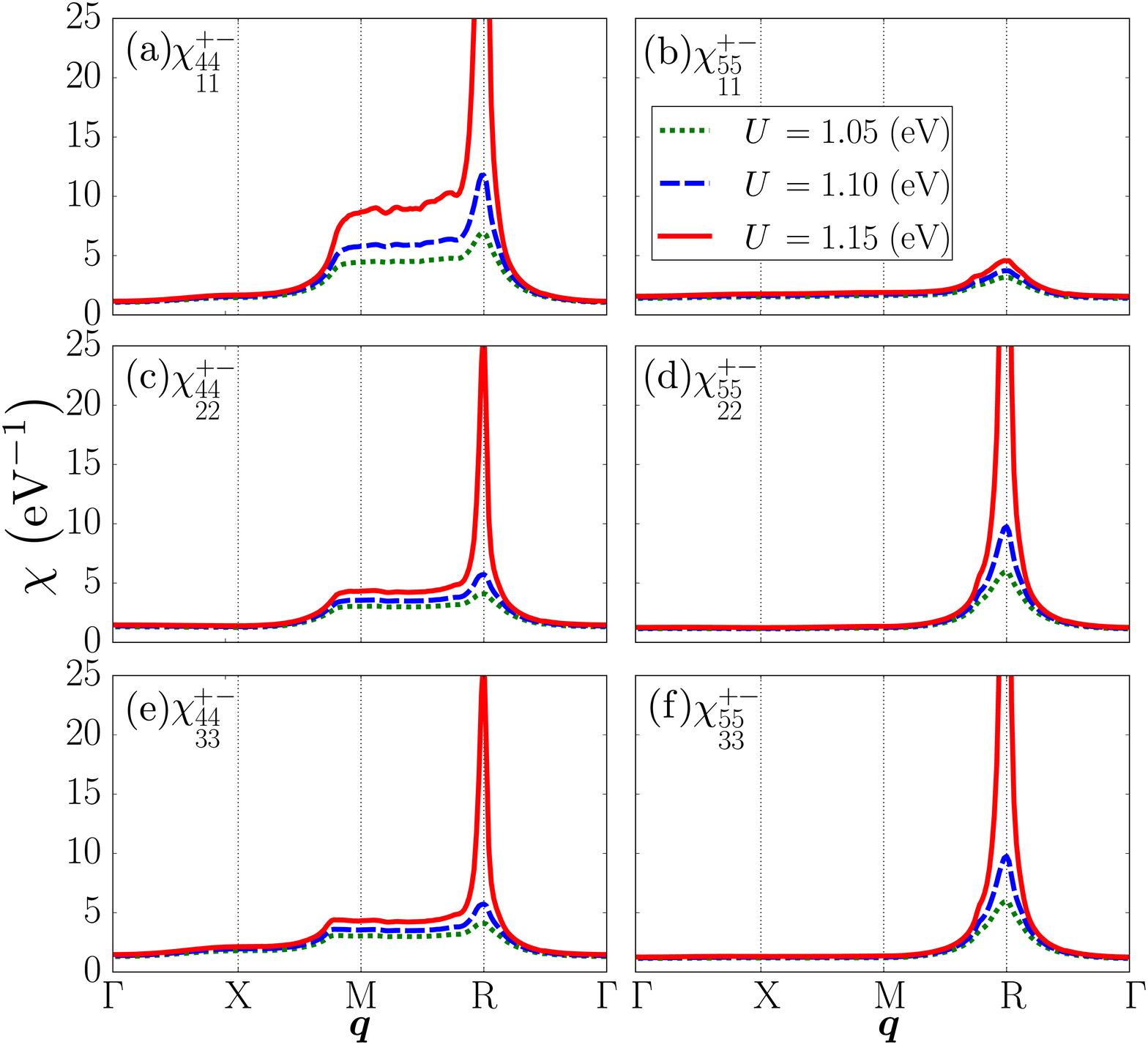}
\caption{(Color online) 
Calculated static susceptibility $\chi^{+-}_{\substack{\mu \mu \\ \nu \nu}}$ in the normal phase.  
The red-solid, blue-dashed, and green-dotted lines are for $U=1.15$, $1.1$, and $1.05$ eV, respectively.
}\label{fig:chit_disorder}
\end{center}
\end{figure}

We calculate the dynamical susceptibilities in the multiorbital RPA, given  by 
\begin{multline}
 \begin{pmatrix}
	\chi^{+-} \\ \chi^{\uparrow \uparrow} \\ \chi^{\downarrow \uparrow}
 \end{pmatrix}
=
 \begin{pmatrix}
	\chi^{+-}_0 \\ \chi^{\uparrow \uparrow}_0 \\ 0
 \end{pmatrix}
\\
+
 \begin{pmatrix}
	\chi^{+-}_0 V^{-+}& 0 & 0\\
	0 & \chi^{\uparrow \uparrow}_0 V^{\uparrow \uparrow} & \chi^{\uparrow \uparrow}_0 V^{\uparrow \downarrow} \\
	0 & \chi^{\downarrow \downarrow}_0 V^{\downarrow \uparrow} & \chi^{\downarrow \downarrow}_0 V^{\downarrow \downarrow}
 \end{pmatrix}
 \begin{pmatrix}
	\chi^{+-} \\ \chi^{\uparrow \uparrow} \\ \chi^{\downarrow \uparrow}
 \end{pmatrix},
\end{multline}
where the matrix product in the orbital basis is given as
\begin{align}
&[\chi_0 V \chi ]_{\substack{\lambda \mu \\ \kappa \nu}}  (\bm{q},\omega) = \notag \\
&\sum_{\kappa', \lambda', \mu', \nu', m}
	{\chi_0}_{\substack{\lambda \mu' \\ \kappa \nu'}} (\bm{q},\bm{q}+m\bm{Q},\omega)
	V_{\substack{\mu' \lambda' \\ \nu' \kappa'}}
	{\chi}_{\substack{\lambda' \mu \\ \kappa' \nu}} (\bm{q}+m\bm{Q},\bm{q},\omega)
\end{align}
with the interaction matrix $V$ listed in Table~\ref{tb:interaction}.

%\section{Results and Discussions} %[[[

%\subsection{Normal phase}

First, let us discuss the spin-triplet excitonic fluctuations in the normal phase.  
Figure~\ref{fig:chit_disorder} shows the $\bm{q}$ dependence of the static susceptibility 
of the excitonic spin-transverse mode $\chi^{+-}_{\substack{\mu\mu \\ \nu \nu}} (\bm{q}, \omega = 0)$ 
calculated in the normal phase, where $\mu$ ($\nu$) is one of the $e_{g}$ ($t_{2g}$) orbitals.  
We find that, at $\bm{q} = (\pi,\pi,\pi)$, the diverging fluctuations with increasing 
$U$ toward $1.15$ eV are observed for all the orbital components except $(\mu, \nu) = (5,1)$.  
This instability toward the EP is caused by the Fermi-surface nesting between the electron 
pockets of the $e_g$ bands located around the $\Gamma$ point of the Brillouin zone and the 
hole pockets of the $t_{2g}$ bands located around the ${\rm R}(\pi,\pi,\pi)$ point of the 
Brillouin zone (see Fig.~\ref{fig:band}).  Thus, the EP transition with the ordering vector 
$\bm{Q} = (\pi,\pi,\pi)$ occurs at the critical value $U_{\mathrm{cr}}=1.15$ eV.

%\subsection{Excitonic phase}

\begin{figure}[th]
\begin{center}
\includegraphics[width=8.5cm]{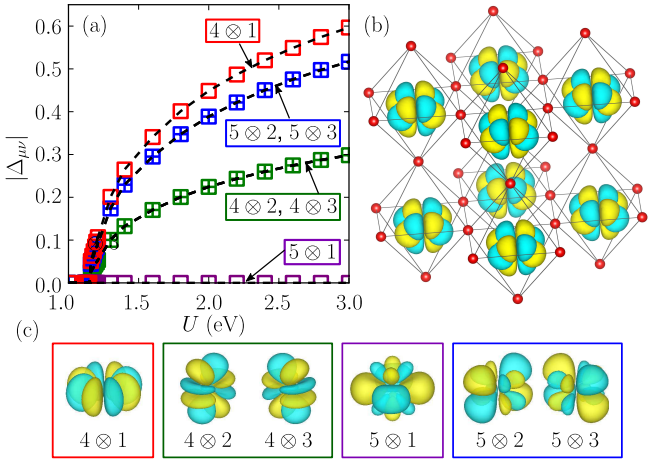}
\caption{(Color online)
(a) Calculated $U$-dependence of excitonic order parameters $\Delta_{\mu,\nu}$.  
(b) Calculated magnetic multipole ordering corresponding to the excitonic order parameter 
at $U=2.5$ eV. 
(c) Schematic representations of the orbital components of the magnetic multipoles.  
}\label{fig:op}
\end{center}
\end{figure}
\begin{figure*}[t]
\begin{center}
\includegraphics[width=17.5cm]{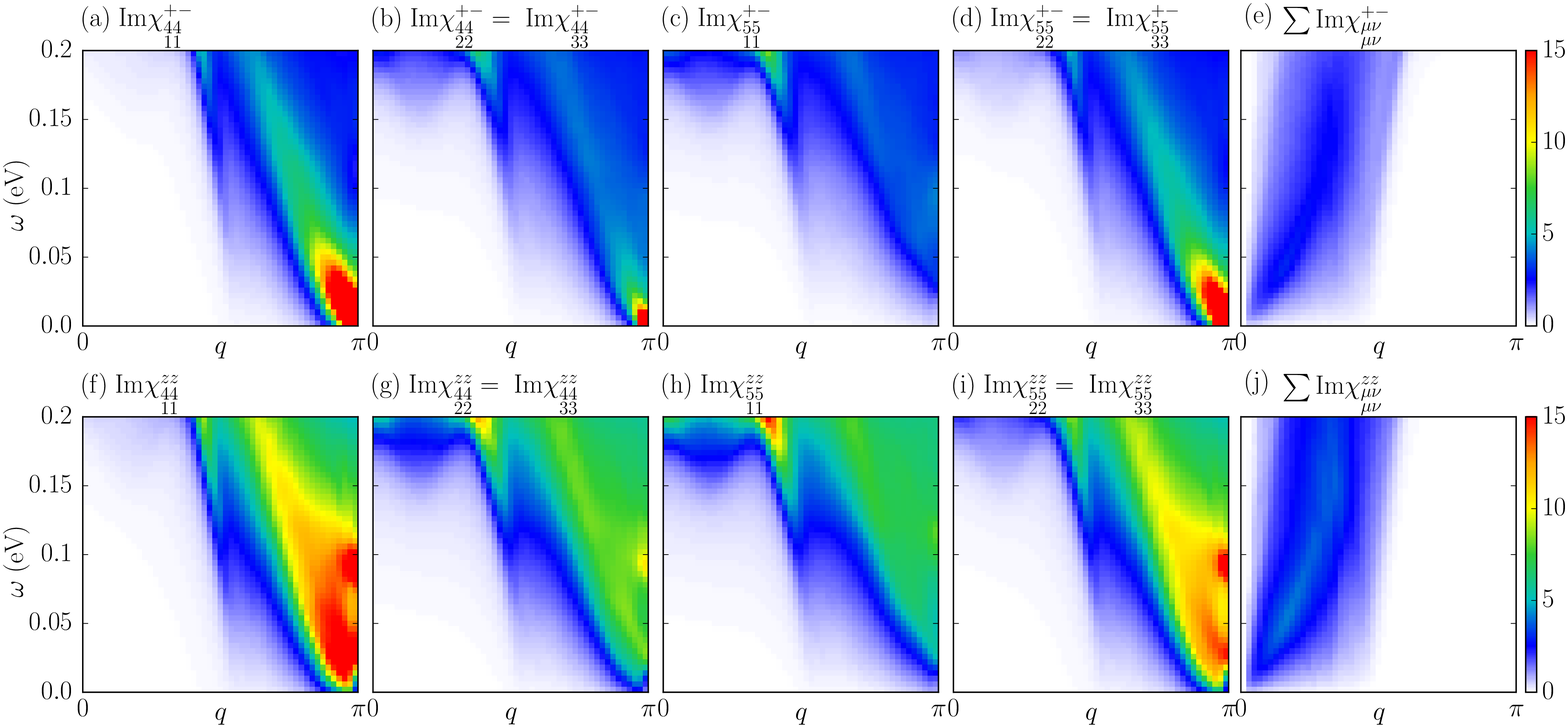}
\caption{(Color online) 
Calculated excitonic spin-transverse (upper panels) and spin-longitudinal (lower panels) 
modes of the dynamical susceptibility in the EP.  
The orbital-diagonal part of the dynamical susceptibility in (e) the spin-transverse and 
(j) spin-longitudinal modes is also shown.  
The path of the wave vector is along the line from $\bm{q} = (0,0,0)$ to $(\pi,\pi,\pi)$.
The interaction strength is assumed to be $U = 1.2$ eV.  
}\label{fig:chi_order}
\end{center}
\end{figure*}

Next, let us solve the mean-field equations to calculate the excitonic order parameter with 
$\bm{Q} = (\pi, \pi, \pi)$.  The obtained orbital components $\Delta_{\mu,\nu}$ are shown in 
Fig.~\ref{fig:op}(a), where we find that all the components $\Delta_{\mu,\nu}$ (except $\Delta_{1,5}$) 
become finite above $U_{\mathrm{cr}}=1.15$ eV.  As $U$ increases, $U'$ and $J$ 
also increase, which enhances $\Delta_{\mu,\nu}$.  Because the excitons are formed 
in a single atom, the excitonic spin polarization leads to the magnetic multipole order 
in real space,\cite{Kunes2014PRB,Kaneko2016PRB} as shown in Fig.~\ref{fig:op}(b).  
The orbital components of the magnetic multipoles formed between the $\mu$ and $\nu$ 
orbitals (indicated as $\mu \otimes \nu$) are shown in Fig.~\ref{fig:op}(c).  
Reflecting the symmetry of the orbitals, the components of the order parameter satisfy the 
relations $\Delta_{4,2}=\Delta_{4,3}=\Delta_{4,1}/\sqrt{3}$, $\Delta_{5,2}=-\Delta_{5,3}=\Delta_{4,1}/2$, 
and $\Delta_{5,1} = 0$, where different combinations of the signs are also possible.  
The last relation indicates that the electrons on the $d_{3z^2 - r^2}$ orbital and holes 
on the $d_{xy}$ orbital do not form pairs.  This is consistent with the result for the 
calculated excitonic susceptibility in the normal phase [see Fig.~\ref{fig:chit_disorder}(b)], 
where no diverging behavior is observed.
%The absence of excitonic pairing in certain orbitals comes from the discrepancy 
%between the number of the conduction bands and the valence bands.
We note that the bands in the EP are not fully gapped at $U<1.8$ eV, keeping the system 
metallic with small Fermi surfaces, which may be consistent with the results of 
experiment.\cite{Hejtmanek2013EPJB}  A full gap opens for larger values of $U>1.8$ eV.  
We also note that the excitonic order remains finite against the change in the filling of 
electrons, e.g., between 5.4 and 6.1 per site at $U=1.2$ eV, which is also consistent with 
experimental results.\cite{Hejtmanek2013EPJB} 

Next, let us discuss the excitation spectra in the EP.  The calculated excitonic spin-transverse 
dynamical susceptibilities $\mathrm{Im}\, \chi^{+-}_{\substack{\mu \mu \\ \nu \nu}}$ are 
shown in Figs.~\ref{fig:chi_order}(a)--\ref{fig:chi_order}(d), where we find the gapless Goldstone 
mode at $\bm{q} = (\pi,\pi,\pi)$ for all the components except for $(\mu, \nu) = (5,1)$ 
[see Fig.~\ref{fig:chi_order}(c)], reflecting the presence/absence of $\Delta_{\mu,\nu}$.  
The velocities of the collective excitations near $\bm{q} = (\pi,\pi,\pi)$ 
are the same for all the components.  
Unlike in the well-known collective mode of the Heisenberg antiferromagnets, 
the excitonic collective mode does not extend to reach the point $\bm{q} = 0$ and $\omega = 0$.  
The gapless collective-mode behavior obtained is consistent with the results of the two-orbital 
models.\cite{Nasu2016PRB, Brydon2009PRB} 
Note that the excitations around $\bm{q} = (\pi/2,\pi/2,\pi/2)$ appearing in all the components 
originate from the particle-hole excitations.  
The calculated excitonic spin-longitudinal dynamical susceptibilities 
$\mathrm{Im}\, \chi^{zz}_{\substack{\mu \mu \\ \nu \nu}}$ 
defined as $\chi^{zz}_{\substack{\lambda \mu \\ \kappa \nu}}=
\sum_{\sigma,\sigma'}\sigma\sigma'\chi^{\sigma\sigma'}_{\substack{\lambda \mu \\ \kappa \nu}}$ 
are also shown in Figs.~\ref{fig:chi_order}(f)--\ref{fig:chi_order}(i), where we find the gapful 
Higgs mode.  The broad excitations that have a gap at $\bm{q} = (\pi,\pi,\pi)$ are clearly seen, 
except for $(\mu, \nu) = (5,1)$ [see Fig.~\ref{fig:chi_order}(h)], where only the particle-hole 
excitations are present.  The spectra around $\bm{q} = (\pi/2,\pi/2,\pi/2)$ appearing in all 
the components are again particle-hole excitations.  
The orbital-diagonal part of the dynamical susceptibilities in the transverse mode, 
$\sum_{\mu,\nu}\mathrm{Im}\, \chi^{+-}_{\substack{\mu \nu \\ \mu \nu}}$, 
and in the longitudinal mode, 
$\sum_{\mu,\nu}\mathrm{Im}\, \chi^{zz}_{\substack{\mu \nu \\ \mu \nu}}$, 
are also shown for comparison in Figs.~\ref{fig:chi_order}(e) and \ref{fig:chi_order}(j), 
respectively, the results of which reflect the metallic nature of the system.  

Finally, let us discuss the possible experimental relevance of our results.  
Inelastic neutron scattering may be a possible experimental technique for observing 
the excitations in the spin degrees of freedom, but may enable one to detect 
only the orbital-diagonal components of the dynamical susceptibility, the intensity of 
which is in proportion to 
$\sum_{\mu, \nu} \mathrm{Im}\, \chi_{\substack{\mu \nu \\ \mu \nu}}^{ss'}$.  
Our corresponding results, which come from the particle-hole transitions, are shown 
in Figs.~\ref{fig:chi_order}(e) and \ref{fig:chi_order}(j).  
Because the dynamical susceptibility related to the orbital-off-diagonal excitonic ordering 
should contain the vertex-nonconserved terms, defined as the terms with $\kappa \neq \lambda$ 
and $\mu \neq \nu$ in Eq.~(\ref{eq:Chi}), we need to seek other quantum-beam sources 
that have the ability to change the orbitals (or orbital angular momentum) in the inelastic 
scattering processes.  To this end, future experimental developments are desired.  

%]]]

%\section{Summary} %[[[

In summary, we derived the effective five-orbital Hubbard model defined on the 
three-dimensional cubic lattice from first principles to describe the electronic states 
of Pr$_{0.5}$Ca$_{0.5}$CoO$_3$ with the cubic perovskite structure.  
Then, we calculated the static susceptibility of the excitonic spin-transverse mode 
in the normal phase using the RPA and found that the diverging excitonic fluctuations 
occur at $\bm{Q} = (\pi,\pi,\pi)$.  We calculated the excitonic ground state in the 
mean-field approximation and found that the magnetic multipole order occurs.  
We also calculated the dynamical susceptibility in the EP to study the excitation spectra 
and found that there appear gapless collective excitations in the excitonic spin-transverse 
mode and gapful collective excitations in the excitonic spin-longitudinal mode.  

%]]]
\medskip
\begin{acknowledgment} 
%[[[
%\acknowledgment
We thank T. Kaneko and S. Miyakoshi for fruitful discussions.
This work was supported in part by Grants-in-Aid for Scientific Research from 
the Japan Society for the Promotion of Science (Nos. 26400349 and 15H06093).
The numerical calculations were carried out on computers at Yukawa Institute for 
Theoretical Physics, Kyoto University, Japan, and Research Center for Computational 
Science, Okazaki, Japan.
\end{acknowledgment}
%]]]

%\appendix %[[[
%\section{}
%]]]

%]]]

\end{document}